\documentclass[12pt,showpacs,preprintnumbers,floatfix]{article}

\pdfoutput=1

\usepackage{graphicx} 
\usepackage{dcolumn} 
\usepackage{bm} 
\usepackage{amsfonts}
\usepackage{amsthm}
\usepackage{amsmath}
\usepackage{amssymb}
\usepackage{epsfig}
\usepackage{color}
\usepackage{textcomp}
\usepackage{hyperref}

\def\be{\begin{equation}}
\def\ee{\end{equation}}
\def\ba{\begin{eqnarray}}
\def\ea{\end{eqnarray}}

\def\bdm{\begin{displaymath}}
\def\edm{\end{displaymath}}

\def\bq{\begin{quote}}
\def\eq{\end{quote}}

 at 10truept

\newcommand{\bea}{\begin{eqnarray}}
\newcommand{\eea}{\end{eqnarray}}

\newcommand{\bi}{\begin{itemize}}
\newcommand{\ei}{\end{itemize}}

\newcommand{\beq}{\begin{equation}}
\newcommand{\eeq}{\end{equation}}
\newcommand{\beqa}{\begin{eqnarray}}
\newcommand{\eeqa}{\end{eqnarray}}

\def\ltap{\ \raise.3ex\hbox{$<$\kern-.75em\lower1ex\hbox{$\sim$}}\ }
\def\gtap{\ \raise.3ex\hbox{$>$\kern-.75em\lower1ex\hbox{$\sim$}}\ }
\def\gl{\ \raise.5ex\hbox{$>$}\kern-.8em\lower.5ex\hbox{$<$}\ }
\def\roughly#1{\raise.3ex\hbox{$#1$\kern-.75em\lower1ex\hbox{$\sim$}}}

\begin{document}

\thispagestyle{empty}
\begin{flushright}
\date{today}
\end{flushright}
\vspace*{1.7cm}
\begin{center}
{\Large \bf Spinodal Backreaction During Inflation and Initial Conditions}

\vspace*{1.2cm} {\large Benoit J. Richard$^{a,}$\footnote{\tt
bjrichard@ucdavis.edu} and McCullen Sandora$^{b,}$\footnote{\tt
sandora@cp3.dias.sdu.dk}}\\
\vspace{.5cm} {\em $^a$Department of Physics, University of
California, Davis, CA 95616}\\
\vspace{.5cm} {\em $^b$CP3-Origins, University of Southern Denmark, Campusvej 55, 5230 Odense M, Denmark}\\

\vspace{2cm} ABSTRACT
\end{center}
We investigate how long wavelength inflationary fluctuations can cause the background field to deviate from classical dynamics.  For generic potentials, we show that, in the Hartree approximation, the long wavelength dynamics can be encapsulated by a two-field model operating in an effective potential. The latter is given by a simple Gaussian integral transformation of the original inflationary potential.  We use this new expression to study backreaction effects in quadratic, hilltop, flattened, and axion monodromy potentials.  We find that the net result of the altered dynamics is to slightly modify the spectral tilt, drastically decrease the tensor-to-scalar ratio, and to effectively smooth over any features of the potential, with the size of these deviations set by the initial value of power in large scale modes and the shape of the potential during the entire evolution.

\vfill \setcounter{page}{0} \setcounter{footnote}{0}
\newpage

\section{Introduction}

Very recent observations \cite{Ade:2015lrj, Planck:2015xua, Ade:2015tva, Ade:2014xna} are on the threshold of making an important distinction about the details of the inflationary regime that gave birth to our universe, by the precise measurement of the parameter $n_s-1$, the slight scale dependence of the primordial perturbations, and the constraint on $r$, the ratio of the power of tensor-to-scalar fluctuations.  The distinction arises because of the slow-roll relations
\be
n_s=-3\frac{V'^2}{V^2}+2\frac{V''}{V}\approx 0.97,\quad r=8\frac{V'^2}{V^2}\lesssim 0.11
\ee
written in term of derivatives of the potential driving inflation, and with observed values given by \cite{Ade:2015lrj, Planck:2015xua, Bennett:2012zja}.  From this we see that the threshold value $r<8/3(1-n_s)\approx 0.08$ implies that the inflaton field has a negative mass squared term, and is thus \emph{tachyonic}.  It is still too early to definitively tell whether this is the case or not, but experiments planned for the immediate future \cite{Bouchet:2011ck,Fraisse:2011xz,Matsumura:2013aja} will be able to distinguish this.
If the inflaton does indeed turn out to be tachyonic, it is worth asking what further consequences this could have for the dynamics of the early universe.  The hallmark feature of tachyonic fields is that long wavelength modes are unstable (over and above the typical instabilities of light fields in an expanding universe).  The possibility that these long wavelengths can have significant enough buildup to heavily influence the dynamics of the background mode was explored in \cite{Cormier:1998nt, Cormier:1999ia}, where the parallel with a similar phenomenon in condensed matter, the $\emph{spinodal regime}$, was drawn.  This is the tachyonic regime, where long wavelength modes experience an instability that leads to exponential growth.  In these and in a subsequent work \cite{Albrecht:2014sea} it was found that backreaction can be significant enough to even drive substantial periods of inflation on steeply tachyonic potentials, where the slow-roll conditions do not hold, such as natural inflation \cite{Freese:1990rb, Adams:1992bn, Freese:2014nla} with subplanckian values of the axion decay constant.  Admittedly, the region of initial conditions suitable for spinodal effects seems quite tuned. However, if the consequences can be so drastic in this special case, it is worth considering whether smaller effects may be present in other models, especially given that data may soon definitively show that the inflaton is tachyonic.  

To this end, we explore whether a spinodal buildup can affect the dynamics of some of the typical models of inflation, and find several general features.  The magnitude of the backreaction effect generically depends on the initial amount of power in the quantum fluctuations.  Accounting for these fluctuations in the energy density does not, in general, cause them to exponentially dilute away, as in the standard scenario where their effect is neglected.  This is actually a consequence of the slow-roll shape of the inflaton potential, which translates into the effective potential governing the strength of the fluctuations being almost equally as flat.  If the inflaton field is tachyonic, exponential buildup of long wavelength modes does occur, but the effects can be seen even in nontachyonic models of inflation if the initial strength is great enough.  We observe several effects:
\begin{itemize}
\item There can be stages where the long wavelength fluctuations dominate the contribution to the power spectrum of density perturbations.  If this is the case, the amplitude is not set by the overall mass scale of the potential, but instead by the initial conditions for the long wavelength modes.  Additionally, this shifts the value of $n_s-1$ by an amount that depends on the potential.  Typically, this is a few percent for large field models, while in small field models the tilt is driven to be exactly scale invariant.  As we will see, this will be enough to alter the predictions of some models to lie within the preferred contour in the $n_s-r$ plane.
\item Though the value of the tilt is only slightly altered, the prediction for the tensor-to-scalar ratio can drastically change if these effects are at play.  This is because the scalar power spectrum can be much larger than the regular contribution, while the tensor power remains unchanged.  This allows for inflation to proceed at a much lower energy scale, which decreases the tensor fluctuations.
\item The spinodal regime will be an inevitable attractor for purely tachyonic models, though approach may be very slow.  In these models, it is possible to differentiate whether inflation lasted only several e-folds longer than what we observe, or if there were a parametrically longer phase preceding this, by noting whether the spinodal regime has been reached.
\item    A final effect is that, if features are present in the potential, the contribution of the fluctuations can serve to wash them out, providing a sort of ``low-pass filter" in which the inflaton is influenced only by the large scale trend of the potential.  This can even make slow-roll inflation possible on otherwise completely unsuitable potentials with many false vacua and steep peaks.
\end{itemize}

The spinodal phenomenon is in actuality nonperturbative, and so to make any headway the Hartree approximation \cite{Chang:1975dt} is employed, which is a self-consistent loop resummation scheme.  The end result of the Hartree approximation is to replace the original model of inflation with a two-field model, where the second field represents the power in the long wavelength modes.  We refer to this second field as the \emph{diakyon}, after the Greek word for fluctuation, and study its influence on the behavior of inflation as a function of its initial value.  Even though the background dynamics is effectively two-field inflation, isocurvature perturbations are not generated, because at the basic level the second field decays exactly the same way as the inflaton during (p)reheating, so the dominant effects of this other field is to alter the trajectory of the inflaton, and to set the value of density perturbations.  What we find is that in spite of the presence of drastically different regimes, where the diakyon field can provide the dominant contribution to the energy density, the perhaps most salient feature of inflation, the production of a nearly scale invariant spectrum of perturbations, remains robust.  We expect the alterations we do observe, a drastic decrease in $r$ and a slight shift in $n_s$, to remain true beyond the Hartree approximation, at the very least at the qualitative level.

Our paper is organized as follows:  we devote section \ref{Hart} to the technical framework.  We begin by using the path integral formalism to  derive the leading quantum corrections to the equations of motion. Afterwards, we discuss the Hartree approximation, how the dynamics is equivalent to replacing long wavelength modes with a second classical field, and show how both can be easily implemented by a simple integral transform of the original potential of the theory.  The section ends with a discussion of density perturbations in this framework. In section \ref{Tart} we apply these results to some specific models of inflation.  We begin with the simplest model of inflation, $m^2\phi^2$, and show that for certain initial values of the diakyon, the tensor-to-scalar ratio can be made much smaller without altering the other predictions of the model.  We go on in section \ref{hillinf} to study the simplest model of a tachyonic field, hilltop inflation, and show that there are several regimes alternative to the standard slow-roll behavior, depending sensitively on the initial value of the diakyon.  In section \ref{flatinf} we look at flattened potentials ($\phi^{2/3}$ and, briefly, Starobinsky), and show that the tachyonic nature of the inflaton necessarily causes deviations from standard inflation if the inflaton started from the eternal regime.   In section \ref{monoinf} we discuss spinodal effects in monodromy inflation.  We find that for small values of the axion decay constant $f$, these serve to taper any oscillations in the spectrum.  We also use this to highlight how spinodal effects can allow for slow-roll inflation even when the potential has many pocket false minima, and discuss this general phenomenon.  We conclude in section \ref{conclusions}.

\section{Hartree Approximation}\label{Hart}
We devote this section to explaining the Hartree approximation as a tool in field theory, and justifying its use in the case at hand.  We provide a simple path integral derivation of the Hartree approximation, which allows us to formally assess the validity of the approximation and, if desired, compute systematic corrections to it.  Though undertaking these tasks is beyond the scope of the present paper, we plan to return to these in future publications.  Afterwards, we distill the Hartree approximation down to a simple integral transform of the potential, allowing for efficient implementation of spinodal effects.

A useful tool in the study of field theories is the background field method.  In this method all fields are divided into background quantities, that may or may not depend on space and time, and quantum fluctuations on top of these.  One is then typically interested in two things: the evolution of the background field, and correlations between perturbations at different spacetime points.  This method is routinely applied in inflationary cosmology, where an inflaton field acquires a time-dependent expectation value driving the expansion of the universe, and the fluctuations source perturbations in the energy density that eventually collapse to form galaxies.  To this end, we assume the dynamics of the early universe is governed by a single field rolling down a potential, and perform the background splitting  
\be
\phi(t,x)=\bar\phi(t)+\psi(t,x).
\ee
This splitting is made unique by the condition that the tadpole of the perturbation vanish,
\be
\langle \psi \rangle=0\Longrightarrow\langle\phi\rangle=\bar\phi.
\ee
To lowest order in $\psi$, this condition implies the classical equations of motion for the background field, and a linear equation determining the fluctuation:  
\bea
\partial_t^2\bar\phi+3H\partial_t\bar\phi+V'(\bar\phi)=0,\nonumber\\
\left(-\Box+V''(\bar\phi)\right)\psi=0.
\label{eom}
\eea

The fluctuations are usually very small, and so then this is a good approximation.  During inflation, however, even though the fluctuations are small, once a mode of a given wavelength gets stretched to larger than the horizon size it freezes to a constant value.  If this occurs for very long time, a large number of modes will exit the horizon and contribute an effective stochastic offset to the equations of motion for the background, that can be encapsulated in the addition of a noise term \cite{Starobinsky:1986fx,Vilenkin:1983xp}, the backreaction of which was recently considered in \cite{Levasseur:2014ska}.  This can lead to drastic inhomogeneities on the largest of scales, and produces divergences if inflation lasts for an exponentially long time (for a review see \cite{Seery:2010kh}), but for the purposes of a single observer measuring local correlators, this only contributes as a shift in the local time coordinate, as set by the value of the background part of the field. In contrast, we are considering the influence of long wavelength modes on the dynamics of these equations that does not manifest itself as a source term.

\begin{figure*}[h]
\centering
\includegraphics[height=3.75cm]{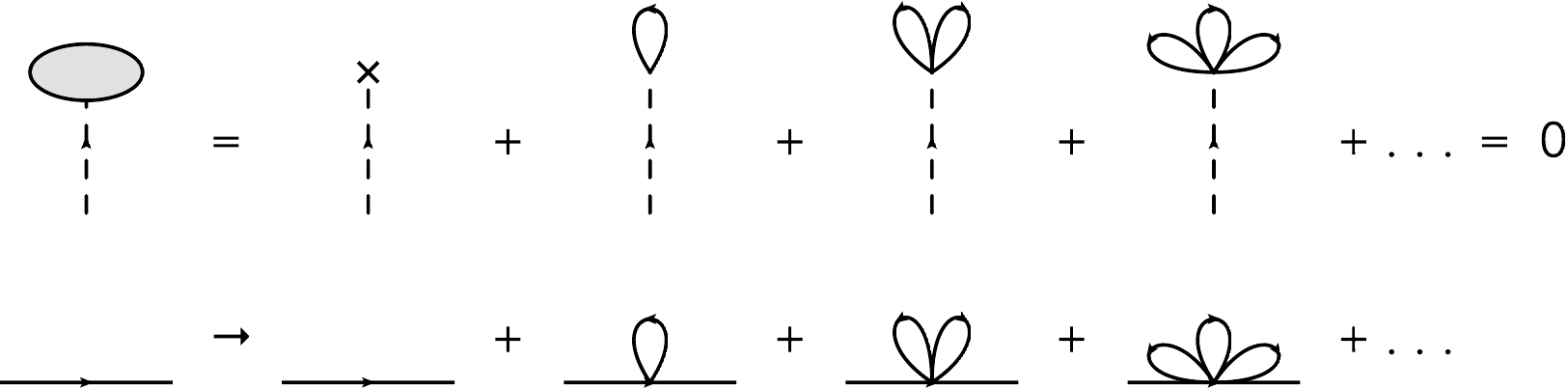}\
\caption{Graphical depiction of the Hartree resummation.  Here the dashed lines correspond to the background field, and the solid lines to the fluctuations.  The top row is the tadpole condition, beginning with a line ending on a $\times$ representing the classical equations of motion.  In the last row, self-consistency demands that once the propagator is substituted with this series, one must iterate the procedure, replacing every propagator in the series itself with the full series of propagators, and so on.  The resulting diagrams lead to this sometimes being called the cactus approximation.}
\label{loops}
\end{figure*}

There is a further reason to expect deviations from the lowest order during inflation if the field is tachyonic.  In this case wavelengths larger than the inverse mass scale of the field will grow exponentially, which can compensate or even overcome the exponential dilution, and can cause significant backreaction on the value of the background field.  In this case the tadpole condition does not reduce to the classical equations of motion, but instead becomes a system depending on an infinite number of the correlators of the fluctuations.  Likewise, the equations determining the values for each of these correlators will depend on a number of the others, as well as the background.  There is no hope of solving this infinite set of equations for an infinite amount of variables analytically.  Fortunately, there is a well-motivated approximation, devised by Hartree, which allows us to truncate this system to a manageable level, and typically yields results that are quite close to the full calculation (see for instance  \cite{Chang:1975dt}).  The approximation consists of treating all propagators as if they were moving through an external medium set by the effect of the propagators themselves.  In this way, we acknowledge that the fluctuations of the fields can yield a large effect on the quantities we are trying to compute.  As shown in Fig \ref{loops}, this method entails replacing all propagators by the sum of insertions of virtual loops, with couplings dictated by the bare Lagrangian.  

Once this replacement has been made, self-consistency demands that  we do the same procedure for propagators flowing through the loops, and so on and so forth, ad infinitum.  It  should not be difficult to convince the reader that this quickly becomes impossibly hard to manage.  Fortunately, there is a more operationally friendly method of enforcing this approximation, as espoused in \cite{Cormier:1999ia}: one need only replace all higher $n$-point functions with products of propagators, multiplying two or fewer powers of the fields.  For example, $\psi^{2n}\rightarrow c_0\langle\psi^2\rangle^n+c_1\langle\psi^2\rangle^{n-1}\psi^2$, where the coefficients are combinatoric factors.  Similarly, for an odd number of the fields $\psi^{2n+1}\rightarrow c_2\langle\psi^2\rangle^n\psi$.  The coefficients will not be important for our purposes, but can be found in \cite{Cormier:1999ia}.  In this way, we have eliminated the need for correlators of much higher powers of the fields, and reduced the system to a set of two equations determining $\bar\phi$ and $\langle\psi^2\rangle$:

\bea
\partial_t^2\bar\phi+3H\partial_t\bar\phi+\sum_{k=0}^\infty\frac{1}{k!}\bar V^{(2k+1)}\left(\frac{\langle\psi^2\rangle}{2}\right)^k=0,\nonumber\\
\left[\Box+\sum_{k=0}^\infty\frac{1}{k!}\bar V^{(2k+2)}\left(\frac{\langle\psi^2\rangle}{2}\right)^k\right]\psi=0.\label{bench}
\eea

These expressions were derived in \cite{Cormier:1998nt} and their dynamics for natural inflation explored in \cite{Cormier:1999ia}.  A few brief comments before we go on to derive this approximation from a path integral point of view:  the Hartree approximation is concerned with allowing the form of propagators to be altered to account for the background shift.  In principle, higher order correlators will also be altered, for example the value of the quartic coupling may receive corrections by inserting interaction loops through different vertices.  This effect is deemed unimportant for the study of two-point functions.  Secondly, it can be seen in Fig. \ref{loops} that only contact insertions are included in the sum, in the sense that the (dressed) internal line originates and terminates at the same place on the propagator, with no further interactions on its journey.  This ignores substructure wherein a second loop might split off from a first and recombine on an entirely different point.  Third, if the Lagrangian has interactions of infinitely high order, such as the ones we will consider throughout this paper, then the Hartree approximation includes arbitrarily high loop order.  This makes it a bona fide resummation scheme, and not a usual loop expansion.  The path integral derivation should dispel any misgiving of giving precedence to extremely high order contact interactions over relatively low order noncontact interactions by showing how this arises naturally from a single insertion of the interaction Hamiltonian.

\subsection{Path Integral Derivation}
Now we turn to a derivation of the Hartree approximation in terms of the path integral.  This will provide the crucial insight we rely on in implementing the Hartree approximation in example potentials.  We start by computing the tadpole of the fluctuation in the path integral formalism
\be
\langle\psi_x\rangle=\int D\psi e^{iS[\bar\phi+\psi]}\psi_x
\ee
The action can then be expanded in powers of the fluctuation $S[\bar\phi+\psi]=\bar S+\bar S^{(1)}\psi+\bar S^{(2)}\psi^2+S_{\text{int}}$, where, since we take the kinetic terms to be quadratic in fields, $S_{\text{int}}=-\int d^4x\sum_{n=3}^\infty\frac{1}{n!}\bar V^{(n)}\psi^n_x$.  Here superscripts represent derivatives with respect to the fields, and barred quantities are evaluated on the background.  Then
\be
\langle\psi_x\rangle=e^{i\bar S}\int D\psi e^{i\bar S^{(1)}\psi+i S_{\text{int}}}\psi_x e^{i \bar S^{(2)}\psi^2}
\ee
So far this expression is exact:  we have just factored out the quadratic part of the action in anticipation of approximating this as a Gaussian integral.  Now we expand the first exponential 
\ba
\langle\psi_x\rangle &\approx& ie^{i\bar S}\int D\psi\left(\bar S^{(1)}\psi+S_{\text{int}}\right)\psi_x e^{i\bar S^{(2)}\psi^2} \nonumber\\
 &=&  ie^{i\bar S}\int D\psi\int d^4y\left(\frac{\delta L}{\delta\phi}\psi_y-\sum_{k=1}^\infty\frac{1}{(2k+1)!}\bar V^{(2k+1)}\psi_y^{2k+1}\right)\psi_x e^{i\bar S^{(2)}\psi^2}
\ea
We have used that the integral of an odd number of $\psi$'s vanishes.  If we were working to lowest order we would also neglect the interaction term, and this would be equivalent to the classical equations of motion.  Now we see the essence of the Hartree approximation:  we insert a single interaction potential in the correlators we wish to compute.  Higher order corrections to this will simply be the neglected terms in the expansion of the exponentials.  We then treat the integrals over the Gaussian as occurring in the free field vacuum,
\be
\int D\psi \psi_x\psi_y e^{i\bar S^{(2)}\psi^2}=\langle\psi_x\psi_y\rangle_0\equiv G_{xy},\quad \langle\psi_x\psi_y^{2k+1}\rangle_0=\frac{(2k+1)!}{2^kk!}G_{xy}G_{yy}^k
\ee
The coincident limit of the Green's function will be divergent as always, so we must regularize to arrive at finite answers.  We make use of results \cite{Boyanovsky:1997mq} showing that the renormalization procedure in this setting does indeed result in replacing bare couplings with renormalized values.  Our investigation will in any case be of the infrared effects of these fields, and so we will not be concerned with the ill-behaved ultraviolet properties of these quantities.  Notice that we have used Wick's theorem to express $n$-point functions in the free vacuum in terms of two-point functions.  This is the crucial fact that makes the Hartree approximation equivalent to summing cactus diagrams.  Enforcing the tadpole condition finally leads to 
\be
0= ie^{i\bar S}\int d^4y\left(\frac{\delta L}{\delta\phi}+\sum_{k=1}^\infty\frac{1}{k!}\bar V^{(2k+1)}\left(\frac{G_{yy}}{2}\right)^k\right)G_{xy}
\ee
This implies that the quantity in parentheses vanishes, equivalent to the first equation in (\ref{bench}).  So far we have only shown half of the equations in the path integral formulation.  Before we discuss the other half we discuss how to replace the coincident limit of the correlators that appear in our expressions with a classical field. 

\subsection{The Diakyon Field}\label{diakyonLegit}

From (\ref{bench}) it can be seen that the background dynamics of the inflaton can be interpreted as if the field were evolving in a two-field model in the transformed potential
\be
V(\bar\phi+\psi)\rightarrow\sum_{k=0}^\infty\frac{1}{k!}\bar V^{(2k)}\left(\frac{\sigma^2}{2}\right)^k.\label{expansh}
\ee
Here the second field $\sigma(y)=\sqrt{G_{yy}}$ represents the power in the fluctuation field.  This second field takes on the role of encapsulating the average amount of power stored in the fluctuations, and has dynamics in its own right.  Hereafter, we refer to it as the \emph{diakyon} after the Greek word for fluctuation, $\delta \iota\alpha\kappa\upsilon\mu\alpha\nu\sigma\eta$. Thus, $\sigma$ appears as a second classical field in our equations of motion, rendering our potential effectively dependent on two fields.

We first must establish the legitimacy of grouping long wavelength modes into a classical field such as the diakyon, as a suitable measure of the effect of tachyonic modes during inflation. Such a two-field interpretation was initially proposed in \cite{Cormier:1998nt}, in which the authors separated the contribution of the correlator into two, based on whether the mode $q$ is inside or outside the horizon. Then
\beq
\sigma = \sqrt{\langle \psi(\vec{x},t)^2\rangle_{q < aH}},\label{root}
\eeq
\noindent corresponding to the diakyon, was defined in such a way that it only comprises of modes for which $q < aH$. Hence, calling upon the equation of motion of the diakyon means setting an increasing dynamical cutoff in $q$-space, $q_{max}$, as a function of time, the horizon $(aH)^{-1}$ decreasing during inflation. A numerical analysis involving a fixed $q_{max}$ and mode equations (see for instance \cite{Albrecht:2014sea}), on the other hand, means considering the presence of both subhorizon and superhorizon modes, regardless of their tachyonic characteristic.

In \cite{Albrecht:2014sea}, the choice was made not to call upon a condensation of all the unstable modes in a classical diakyon-like field, in favor of a cutoff in $q$-space. Using mode equations, however, one needs to make choices in order to comply with numerical limitations while still accounting for most of the effect due to the presence of tachyonic modes. One choice corresponds to $q_{max}$. Ideally, one should let $q$ run to infinity. The validity of effective field theory (EFT) allows us to have a finite cutoff in $q$. However, the higher the $q_{max}$ the greater the total number of equations in the system, the latter itself depending also on the step size in $q$ of our numerical analysis. As time goes on, more and more modes will exit the horizon and have some participation in the overall spinodal effect.  Hence, most of the impact attributed to the spinodal instabilities happens during early times, as a result of the long wavelength modes crossing the horizon \cite{Cormier:1998nt}. At very early times, however, $aH$ is vanishingly low, which in turn implies that the lowest few $q$-modes are responsible for the major part of the spinodal effect. Therefore, for such times, requiring a high $q_{max}$, while seemingly more precise, may also be unnecessary. Encompassing the lowest $q$-modes into a classical field thus looks efficient to incorporate most of the consequences resulting from tachyonic instabilities.

Our goal in this paper is not to give a very precise quantitative picture but a more qualitative one. After all, we are always limited by the accuracy of numerical computations. In \cite{Boyanovsky:1997xt} a lower bound on the total number (beyond 360 depending on the mass of the field) of e-folds necessary to study spinodal effects after grouping the mean field and quantum fluctuations into one classical field, was given, in the large N approximation. An argument was also made as to why such a classicalization is valid after only a few e-folds. The observation that the long wavelength modes are responsible for most of the spinodal effect \cite{Cormier:1998nt}, implies that grouping the modes which exit the horizon the earliest into one classical mode may be a valid approximation, especially if the total number of e-folds in the model is large.  Allowing initial conditions that provide enough inflationary e-folds, and considering times 50 to 60 e-folds before the end of inflation should be sufficient to give a consistent qualitative picture of the effects of spinodal instabilities on inflation.

One should also address the choice of initial conditions. In refs. \cite{Cormier:1998nt} and \cite{Boyanovsky:1997xt} an effective initial condition for the diakyon was given, 
\beq
\sigma_0 = \frac{H}{2\pi}, \label{dstemp}
\eeq

\noindent corresponding to a zero-temperature vacuum initial state. A vacuum state is a simple but nonetheless valid choice of initial state. Given the previous discussion, how much the spinodal instabilities manifest themselves depends on the initial value for the diakyon field compared to the inflaton. Thus, the number of initial conditions is the same whether we choose to call upon the mode equations, or a two-field picture. In the former framework, one needs to choose $q_{max}$ and $\phi_0$ while in the latter $\sigma_0$ and $\phi_0$ need to be set. Another parameter that needs to be fixed in both scenarios is the scale of inflation.  If the scale of inflation is low, this allows the amplitude of $\langle \psi^2 \rangle$ to be large enough that it has a significant contribution to the overall dynamics of our system, while still maintaining the correct predictions for the other inflationary observables. 

Having discussed the regime of validity of our classical two-field framework, we can now proceed to discussing how to transform our potential from a function only involving one field, to one suitable within a two-field construction. 


\subsection{The Hartree Transform}


Eq. \eqref{expansh} is a cumbersome formula that involves Taylor expanding the potential of the field, then replacing each factorial coefficient by a different number.  This is not only laborsome, but can quickly become impossible, even for some of the relatively simple potentials we consider here. Additionally, the convergence properties are worsened by this substitution; even for the simple example of noninteger monomials this transformed expression fails to converge.

Fortunately, the expression (\ref{expansh}) can be reproduced by a very simple integral transform of the potential, circumventing the need to ever Taylor expand in the first place.  We define the $\emph{Hartree Transform}$\footnote{this is actually known as a Weierstrass transform, itself just a convolution against a Gaussian} of a function to be 
\be
\mathbb{H}[V](\bar\phi,\sigma)\equiv V^\mathbb{H}(\bar\phi,\sigma)=\int_{\mathbb{R}} d\psi\frac{1}{\sqrt{2\pi}\sigma}e^{-\frac{\psi^2}{2\sigma^2}}V(\bar\phi+\psi)\label{hart}
\ee
It can be easily seen that the result of the Gaussian integration enforces the correct alteration of the coefficients to reproduce the Taylor-expanded result (\ref{expansh}).  This transformation also has a very natural interpretation from the path integral perspective, as it exactly coincides with the approximation we made there to insert a single power of the interaction potential in the free-field vacuum (the quadratic term, though not a part of the interaction potential, automatically assembles with the rest of the terms to yield the unsplit potential).  In fact, we have the general formula: $\mathbb{H}[f(\psi)]=\langle f(\psi)\rangle_0$, and can make use of the fact that derivatives with respect to background quantities can be pulled out of the brackets to directly rewrite all results in terms of the Hartree transform.  The integration calls for the function to be defined everywhere on the real axis, which may not be part of the dynamically accessible regime, as in fractional powers of fields or logarithms.  In order to ensure the reality of the result, we choose the nonunique but physically reasonable prescription that the function vanishes in inaccessible domains.  

Let us interpret this expression to get an intuitive understanding of what the Hartree approximation actually does.  Here we see that we are smearing over the fluctuations with a Gaussian filter, with a variance that is allowed to depend on time as determined by the new equations of motion.  This will impose that the value of the new field will be given exactly by the variance of the fluctuations, and will treat the evolution of the background field as occurring in the ``homogenized" replacement field, the diakyon.  Statistically speaking, this will give the same results as if we were to calculate the background and mode functions of the fluctuations, and compute the two-point corrleator for those (to the extent that the approximation holds).  Additionally, in the limit that the fluctuations do not have a strong influence on the background mode, the field $\sigma$ approaches zero, and
\be
\mathbb{H}[V](\bar\phi,0)=\int_{\mathbb{R}}d\psi\delta(\psi)V(\bar\phi+\psi)=V(\bar\phi).
\ee
The distributional nature of the small $\sigma$ limit will make this formula cumbersome for some purposes, but the expansion (\ref{expansh}) is quite easy to handle.  For instance, we see that the mass of the diakyon at 0 is $m^2_\sigma=\bar V_{\phi\phi}$, which encapsulates the fact that the strength of fluctuations will only grow if the inflaton is tachyonic.  In this sense, applying the Hartree approximation will automatically tell us if the quantum corrections to the equations of motion are important or not.  If we find that the diakyon is driven to $0$ (or even the small value $H/(2\pi)$, as the case would be in de Sitter space) for any initial value on a given potential, we can conclude that the usual lowest order approximation is sufficient, while if we uncover that the diakyon is not pushed to small values we conclude that these effects are important.  For the slow-roll potentials we consider, the mass of the inflaton is much smaller than the Hubble rate, which implies that the potential in the fluctuation direction is similarly flat.  This means that even for positive mass squared the field is not driven to its standard value $\sigma\sim H/(2\pi)$ very quickly; in fact, for chaotic initial conditions we would expect the two values of the fields to be comparable, meaning that the fluctuations would not necessarily reach the attractor value until near the end of inflation.  This is in stark contrast to the scenario in which the backreaction is not taken into account, where any initial power is exponentially diluted away and the field equilibrates to the de Sitter temperature.  This translates the question of the importance of quantum corrections to the dynamics to a problem of initial conditions for the perturbations.  If we begin inflation on a potential that is otherwise suitable for slow-roll, but the initial value of the diakyon is very large, it will continue to play an important role throughout the evolution.  In addition, if we envision that inflation started with a fast-roll phase \cite{Linde:2001ae}, in which the inflaton has a large tachyonic mass, then the diakyon quickly gets driven \emph{away} from its standard value, and so we would expect to find that the slow-roll evolution proceeds with a much stronger amount of fluctuations, regardless of when in the inflationary history the fast-roll event took place.

We end this discussion by noting the important identity 
\beq
\boxed{\sigma\partial_\phi^2V^\mathbb{H}(\phi,\sigma)=\partial_\sigma V^\mathbb{H}(\phi,\sigma)}\label{heat}
\eeq
which holds at every point on the potential.  This can be verified directly, but can also be arrived at by the observation that if we formally replace $\sigma^2=t$, $\phi=x$ then this is the heat equation, and (\ref{hart}) resembles the heat kernel.  This equation will be an important tool for studying the behavior of the system, and we make use of it many times in the remainder of this paper.  For example, we can use it to find an expression for the mass of the diakyon that is valid for any value of the fields:
\beq
\partial_\sigma^2V^\mathbb{H}=\partial_\phi^2V^\mathbb{H}+\sigma^2\partial_\phi^4V^\mathbb{H}.
\eeq
This recovers that the masses of the two fields are equal for small $\sigma$, but is an exact expression that informs us that the mass of the fluctuation can either be significantly larger or smaller than the inflaton for large field values.

\subsection{Diakyon Dynamics}
We have shown that the background dynamics can be encapsulated by adding an additional diakyon field representing the strength of the fluctuations and replacing the potential by its Hartree transform.  We now show that these replacements perfectly capture the behavior of the fluctuations as well.  In addition to the background dynamics, the equations governing the fluctuations can be derived in the path integral formalism.  To do this, it is convenient to briefly introduce the two-particle irreducible formalism \cite{Luttinger:1960ua,Baym:1961zz,Cornwall:1974vz}, since this is specially designed to study the behavior of the fluctuation spectrum in the regime where quantum effects can become important.  This simply entails adding a bilocal source for the fluctuations in addition to the usual source introduced when we are only interested in one-particle irreducible questions, as demonstrated by
\beq
Z[j_x,k_{xy}]=\int D\psi e^{iS[\bar\phi+\psi]+i\int dxj_x\psi_x-\frac i2\int dxdyk_{xy}\psi_x\psi_y}.
\eeq
The effective action is then defined as the Legendre transform of the logarithm of the previous quantity with respect to both $j_x$ and $k_{xy}$.  Enforcing that the derivative of $Z[j_x,k_{xy}]$ with respect to $j_x$ vanishes yields the tadpole condition $\langle\psi\rangle=0$ that we explored above.  In addition, we impose the gap equation, namely that the derivative with respect to $k_{xy}$ also vanishes.  This will yield a nonlinear equation that dictates the spectrum of fluctuations.  In the full theory, this will again involve correlators of  arbitrarily many fields, and will be unsolvable.  We can employ the same Hartree approximation to truncate the series to a finite subsystem, which yields
\beq
\Big(\Box_x+ V_{\phi\phi}^{\mathbb{H}}\left(\bar\phi,G_{xx}^{1/2}\right)\Big)G_{xy}=i\delta(x-y).
\eeq
If the two-point function is small, this leads to the lowest order result that the two-point correlator is just the Green's function of the second variation of the action.  This standard result can be significantly altered in regimes where the two-point function is much larger, as explored in \cite{Riotto:2008mv}, where it was shown that self interactions serve to taper secular growth of correlators of massless fields (also see \cite{Youssef:2013by}, where a refinement on the Hartree approximation was recently employed).  Since the Green's function is just the time ordered product of the mode functions, $G_{xy}=\mathcal{T}[\psi_x\psi_y]$, this equation is equivalent to the second equation in (\ref{bench}).  

To show that the dynamics of these fluctuations can be replaced with the classical evolution of the diakyon field, we simply use (\ref{heat}) to rewrite (\ref{bench}) as 
\beq
\Box\sigma+\partial_\sigma V^\mathbb{H}(\phi,\sigma)=0.
\eeq
With this, we have shown that the entire background dynamics can be reproduced as two classical fields in the altered potential $V^\mathbb{H}(\phi,\sigma)$, and have completely recovered the dynamics laid out in \cite{Cormier:1998nt,Cormier:1999ia}\footnote{There is also the Friedmann equation determining $H$ in terms of the background and fluctuations.  This, and any other equation for fields not directly involving $\psi$, are trivially reproduced by the Hartree transform.}. Since we have justified to call upon $\sigma$ to encompass the effect of fluctuations, we will omit the bar in $\bar{\phi}$ and refer to $\phi$ as the inflaton, in the rest of the paper.

\subsection{Density Perturbations}
Though the background behavior is encapsulated as a simple two-field model, the diakyon is in actuality a composite of the inflaton field, and so we would expect fluctuations in the energy density to be different from the usual two-field formula.  The expression was derived in \cite{Cormier:1999ia} by calculating the perturbation of the time-time component of the stress-energy tensor \emph{before} resumming the modes into the mean field $\sigma$.  This is the most convincing procedure for obtaining the spectrum, but here we content ourselves with reproducing their results by a simpler method, where we employ immediate use of the diakyon approximation. Then
\beq
P_k=\frac16\frac{\langle \left(V(\phi+\psi)-V(\phi)\right)^2\rangle_0}{\dot H^2}=\left(\frac{V^\mathbb{H}}{(V^\mathbb{H}_\phi)^2+(V^\mathbb{H}_\sigma)^2}\right)^2\Big(\mathbb{H}[V^2]-2V\mathbb{H}[V]+V^2\Big)\label{spectrum}
\eeq
This expression is fully general.  In the limit of small fluctuations, which is the case for instance if they are set by the typical de Sitter fluctuations, the perturbed potential can be Taylor expanded and the expression becomes
\beq
P_k\rightarrow\left(\frac{V^\mathbb{H}}{(V^\mathbb{H}_\phi)^2+(V^\mathbb{H}_\sigma)^2}\right)^2\langle V_\phi^2\psi^2\rangle_0.
\eeq
Using an identity similar to (\ref{heat}) this can be rewritten as
\beq
P_k\rightarrow\left(\frac{V^\mathbb{H}}{(V^\mathbb{H}_\phi)^2+(V^\mathbb{H}_\sigma)^2}\right)^2(\sigma^2+\sigma^4\partial_\phi^2)\mathbb{H}[V_\phi^2],
\eeq
which, upon taking the limit $\sigma\rightarrow 0$ agrees with the result derived in \cite{Cormier:1999ia}.

The $\sigma\rightarrow 0$ limit in this expression appears somewhat pathological, due to the apparent vanishing of the power spectrum.  In this limit the standard scenario is in fact recovered, as the power of the large-wavelength modes can never go below the de Sitter value, meaning that $\sigma\rightarrow H/(2\pi)$, where it tracks the value of the potential.  Operationally, we make the replacement $\sigma^2\rightarrow\sigma^2+V/(12\pi^2)$ in all our expressions, as justified by the definition in (\ref{root}).  If we treat $\sigma^2/M_p^2$ as a small parameter, all but the leading dependence can be neglected in the expression for the power.  In this case the Hartree transform becomes the trivial integration against a delta function, and the expression reduces to 
\beq
P_k\rightarrow \frac{V^2}{V_\phi^2}\left(\sigma^2+\frac{V}{12\pi^2}\right)\label{spectro}
\eeq
The interpretation of this limit is clear:  the $V^2/V_\phi^2$ prefactor is the usual quantity that makes the expression gauge invariant, and the term in parenthesis is the expected value of the two-point function.  If the initial power on large scales is negligible, then this is just given by the de Sitter temperature, whereas if the power on large scales is large it is given by the initial value.  What is nontrivial about this expression is the fact that now we are working in a consistent framework that dictates the behavior of large scale power for any given value after evolution.  One crucial feature to note now is that in the limit where the diakyon dominates the power, the power spectrum does not depend on the overall scale of the potential.  This is in great contrast to the usual case, where the overall scale of the inflaton potential is commonly set to give the observed normalization. Here this condition sets the initial value of the diakyon field.  This translates the usual practice of trying to understand the amplitude of perturbations in terms of the microphysics dictating the precise potential to a question of why the diakyon took the value that it did.  This also allows a set potential to yield a range of values for the fluctuations, making the observed value a potentially environmental quantity.  

To match the value we see, we arrive at
\beq
\frac{\sigma}{M_p}=\sqrt{\epsilon P_k}\ll 1,
\eeq
so that our expansion in small $\sigma/M_p$ is indeed justified.  Note, however, that even though the diakyon is small compared to the Planck scale, we are still free to be in the regime where it can be larger than the de Sitter temperature.  From these considerations, we expect that the diakyon will have its greatest effect in small field models of inflation, where the inflaton is much smaller than the Planck scale as well.

We end this section by discussing the measurable quantities related to the power spectrum.  Still in the limit $\sigma\ll M_p$, we arrive at
\begin{equation}
n_s-1=-\epsilon\left(4+\frac{2}{1+y}\right)+2\eta\frac{1}{1+y}\rightarrow \left\{
 \begin{array}{rl}
  -6\epsilon+2\eta & y\rightarrow 0\\
   -4\epsilon & y\rightarrow \infty
 \end{array} \right.
\end{equation}
Where we have introduced $y=12\pi^2\sigma^2/V$ as the dimensionless measure of fluctuations on large scales.  In the large diakyon limit, the contribution from $\eta$ completely drops out!  This tells that for small field models, where $\epsilon\ll1$ \cite{Lyth:1996im}, the spectrum becomes practically scale invariant in this limit.  For large field models, whether the tilt increases or decreases depends on whether $\epsilon$ or $\eta$ is larger.  In the eventuality $\epsilon=\eta$, the prediction for the tilt remains constant throughout the transition to the spinodal regime.  This is the case for quadratic potentials, and we explicitly verify this in section \ref{m2phi2}.

Similarly, we can derive an expression for the tensor-to-scalar ratio, where we simply use the fact that the value of tensor fluctuations are still set by the (Hartree transformed) potential
\begin{equation}
r=\frac{16\epsilon}{1+y}\rightarrow \left\{
 \begin{array}{rl}
  16\epsilon & y\rightarrow 0\\
   0 & y\rightarrow \infty
 \end{array} \right.
\end{equation}
which displays the alluded tendency to vanish for large values of the diakyon.  Taken in combination, these two observables indicate that a large value of the diakyon tends to push predictions towards the bottom right corner of the $n_s-r$ plot.  Simple application of this formula presumes that the values of $\epsilon$ and $\eta$ remain unchanged no matter the value of the diakyon, in particular that the value of the inflaton field 50 or 60 e-folds before the end of inflation is unaltered.  This is certainly true for small enough values of the diakyon, but away from this limit the field trajectories can be solved numerically.

From these expressions we notice a general tendency of spinodal effects to push inflationary predictions down and to the right in the $n_s-r$ plane.  This motivates these effects to be considered especially in models that traditionally lie to the left of the Planck contour, in the hopes that corrections might bring them back into agreement with observation.  We choose to illustrate this in the simplest scenario of hilltop inflation in section \ref{hillinf}, but some other potentials this could potentially be relevant for include inflection-point inflation, Coleman-Weinberg inflation, double well inflation, and pseudonatural inflation \cite{Martin:2014vha}.

\section{Predictions for Specific Potentials}\label{Tart}
\begin{figure*}[h]
\centering
\includegraphics[height=9cm]{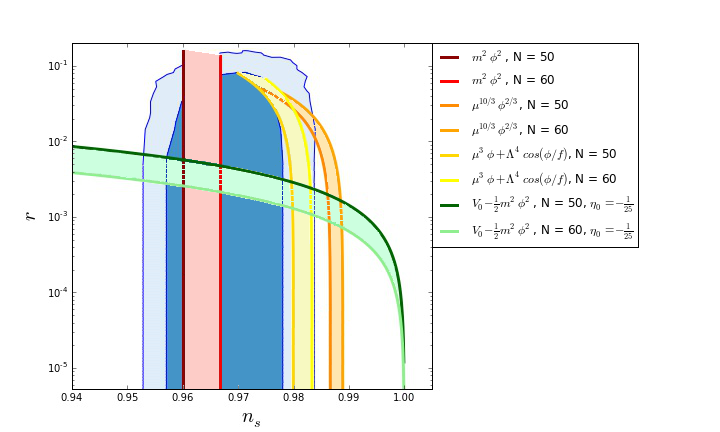}
\caption{The $n_s-r$ plot for the models we consider in this paper, along with the Planck $1\sigma$ and $2\sigma$ contours.  Allowing for strong backreaction can serve to place both $m^2\phi^2$ and hilltop inflation  predictions firmly in the preferred region, and kicks the flattened model out.  This highlights the generic trend that tensor modes become more negligible the stronger the back reaction is.}
\label{master}
\end{figure*}

We now apply our findings in the previous section to some specific potentials, to see what effects the backreaction might have.  We begin with the simplest model of chaotic inflation, a single field in a quadratic potential, in section \ref{m2phi2}.  The Hartree transform of this model consists of two noninteracting fields with equal mass.  The only influence of the diakyon field in this setup is through the fluctuations, and can serve to drastically decrease the tensor-to-scalar ratio, while leaving the spectral tilt fixed.  We then turn to the simplest tachyonic model, quadratic hilltop inflation, in section \ref{hillinf}.  We show in this model that the tilt is driven to be exactly scale invariant and the tensor perturbations unobservable, but that it takes typically $\sim100$ e-folds to reach this regime, providing a window agreeing with the observed values that could possibly encompass cosmic microwave background (CMB) scales.   A tachyonic large field model is considered in section \ref{flatinf} where we consider a flattened potential.  This model eventually attains a spinodal regime for practically all initial conditions, allowing us to infer the difference between inflation descending from the eternal regime and inflation that lasted only marginally longer than the observed amount. These findings are summarized in Fig.\ref{master}.
As a final application, we study monodromy models, where we see that backreaction effects serve to dampen the oscillations of the potential for small enough values of the axion decay constant $f$.  This can even provide a route for the inflaton to escape false minima, allowing for slow-roll even in potentials that would not normally allow it.

Before considering specific models, we first exhibit a useful general result, applicable if the value of the diakyon remains small compared to the Planck scale.  We remind the reader that we only consider this regime anyway, since otherwise (\ref{spectrum}) would imply that density perturbations are well above unity.  In this regime, the field dynamics, in the slow-roll regime, becomes simply
\beq
V\frac{d\phi}{dN}=-M_p^2V_\phi,\quad V\frac{d\sigma}{dN}=-M_p^2V_\sigma.
\eeq
Here $N$ is the number of e-folds.  Several identities can be used to rewrite the diakyon equation in a simpler form.  For instance, the heat equation identity (\ref{heat}) can be used to express this as a linear equation in $\sigma$
\beq
V\frac{d\sigma}{dN}=-M_p^2V_{\phi\phi}\sigma,
\eeq
and to lowest order $V$ is the original, $\sigma$ independent potential.  Additionally, if the inflaton evolution is monotonic, we can express this as $\sigma(\phi(N))$ to arrive at
\beq
\frac{d\sigma}{d\phi}=\frac{V_{\phi\phi}}{V_{\phi}}\sigma,\label{lindiak}
\eeq
which has the solution
\beq
\sigma(N)=\sigma_0e^{\int_{\phi_0}^{\phi(N)}d\phi\frac{V_{\phi\phi}}{V_{\phi}}}=\sigma_0e^{\int_{V_{\phi_0}}^{V_{\phi(N)}} d(\log{V_\phi})}=\sigma_0\frac{V_{\phi(N)}}{V_{\phi_0}}.\label{integ}
\eeq
This allows us to relate the total amount of buildup of large scale power to properties of the potential evaluated at specific points.  It is remarkable that this equation is integrable, and so we reflect on why indeed this is the case.  In general, one expects an exponential buildup of modes for tachyonic masses to be given by the integral of (\ref{lindiak}), and modes to subside during periods of positive mass squared.  If this were all the information we had about the system, it would not be integrable, as there is no way to relate the amount of time spent with any given value of the mass to the endpoints of the evolution.  However, since the dynamics is tied to the evolution of the inflaton, periods of large mass squared are passed through more quickly than periods of small mass squared, and the entire evolution can be encapsulated in the difference between beginning and end points.  We note the pathological limits of this expression, where the slope of the potential vanishes:  if it vanishes initially, inflation lasts an infinitely long time, and the diakyon attains infinite values.  Similarly, if we evolve this to the point where the slope vanishes at the end, the diakyon subsides completely.

\subsection{$m^2\phi^2$}\label{m2phi2}
We begin with a study of the simplest model of inflation, where the potential is just a quadratic function of a single inflaton field to show that, even though the field is not tachyonic, the fluctuations can have a large effect on the background dynamics.  We stress that large values of the diakyon are not evolved to dynamically, so in order for effects to be important in this scenario the diakyon must start out very large.  Thus, we cannot claim the effects we find to be generic predictions for the model as we can in the following sections, but we include this case anyway because its simplicity enables a clear demonstration of how effects can come into play.  

This potential is simple enough that the evolution of the fields can be solved exactly in the slow-roll approximation.  The Hartree transform yields a two-field potential 
\beq
V^{\mathbb{H}}_{\text{chaotic}}(\phi,\sigma)=\frac12m^2\left(\phi^2+\sigma^2\right),
\eeq
which is simply two noninteracting fields with the same mass.  Note that for the background dynamics it is possible to exploit the $O(2)$ symmetry to arrange that only one field direction has an initial value.  The perturbation spectrum, however, does not respect this symmetry, and so we prefer to work with the physical fields.

Using the slow-roll equations
\beq
(\phi^2+\sigma^2)\frac{d\phi}{dN}=-2M_p^2\phi,\quad (\phi^2+\sigma^2)\frac{d\sigma}{dN}=-2M_p^2\sigma,
\eeq
we can solve these exactly
\beq
\phi(N)=\phi_0\sqrt{1-\frac{4M_p^2N}{\phi_0^2+\sigma_0^2}},\quad \sigma(N)=\sigma_0\sqrt{1-\frac{4M_p^2N}{\phi_0^2+\sigma_0^2}}.
\eeq
Both fields reach $0$ at precisely the same time, $\sigma$ being just a rescaled copy of $\phi$.  If we make the approximation that this is the point at which inflation ends, then we can use equation (\ref{spectrum}) to find
\beq
N_{\text{CMB}}=\frac{\phi_0^2+\sigma_0^2}{4M_p^2},\quad P_k= N_{\text{CMB}}\frac{\sigma^2}{M_p^2}+\mathcal{O}\left(\frac{\sigma}{M_P}\right)^4 ,\quad n_s-1=-\frac2{N_{\text{CMB}}},\quad r=\frac{4}{3\pi^2}\frac{m^2}{\sigma_0^2}.
\eeq
In contrast with the usual predictions, now the amplitude of the power spectrum is set by the value of the diakyon field instead of the mass scale of the potential.  If we insisted on the standard value, $\sigma=H/(2\pi)$, the usual results would be recovered, and the value of the mass would be needed to be set to reproduce the observed spectrum.  If the diakyon is large, however, normalization of the power spectrum implies 
\beq
\sigma_0\approx\sqrt{P_k/N_{\text{CMB}}}M_p\approx 6\times10^{-6}M_p\ll15M_p\approx\phi_0.
\eeq
This indicates that the initial value of the inflaton is much larger than the initial value of the diakyon, but the diakyon still has a much larger value than the usual case for generic $m^2$.  Evidently, the tilt is still only sensitive to the total number of e-folds, and so its standard predictions remain unaltered.  The only parameter that depends on the mass scale here is the tensor-to-scalar ratio, which now decreases when $m^2$ takes small values.  This has the consequence that the predictions of this model can be brought into agreement with the preferred values of these parameters, if the mass scale is sufficiently small.  The origin of this result resides in the fact that, if backreaction is large, fluctuations are no longer tied to the expansion rate, but can now be much larger.  Since the strength of the scalar fluctuations is dictated by the value of $\sigma$, whereas the tensor fluctuations are still prescribed by the overall energy density, the scale of the potential can be lowered arbitrarily, while still maintaining the observed power spectrum.

\subsection{Hilltop Inflation}\label{hillinf}
While the quartic hilltop model fits with the measured values of inflationary parameters well enough to merit inclusion on the Planck 2015 $n_s-r$ plot \cite{Ade:2015lrj}, the simplest model, quadratic hilltop inflation, predicts too small an $n_s$.  As such, it is the ideal case to study.  We find already in this simple scenario that there are many different regimes, corresponding to the initial value of the diakyon.

The transformed potential in this case is
\beq
V_{\text{hilltop}}^\mathbb{H}(\phi,\sigma)=V_0-\frac12m^2(\phi^2+\sigma^2)
\eeq
and, similar to the previous case of $m^2\phi^2$, near the top of the hill the two fields behave independently of each other, each experiencing exponential growth:
\beq
\phi(N)=\phi_0e^{|\eta_0|N},\quad \sigma(N)=\sigma_0e^{|\eta_0|N},
\eeq
where $\eta_0=-m^2M_p^2/V_0$ is the slow-roll parameter at the top of the hill.  For simplicity, we consider the predictions for the case when inflation ends at the point where the potential becomes negative.  In actuality, we expect the form of the potential to be altered well before this, but this simplistic setting will be enough to illustrate the effects of the diakyon field.  In this scenario the vacuum energy can be approximated by the constant term in the potential, with corrections becoming important only in the last e-fold or so of the evolution.  The value of the inflaton at CMB scales is set to be $\phi_{\text{CMB}}=\sqrt{2/\eta_0}e^{-|\eta_0|N_{\text{CMB}}}$. The initial value of the diakyon must be less than this, or else it would spoil inflation before the required amount of e-folds has elapsed.  

\begin{figure*}[h]
\centering
\includegraphics[height=6.25cm]{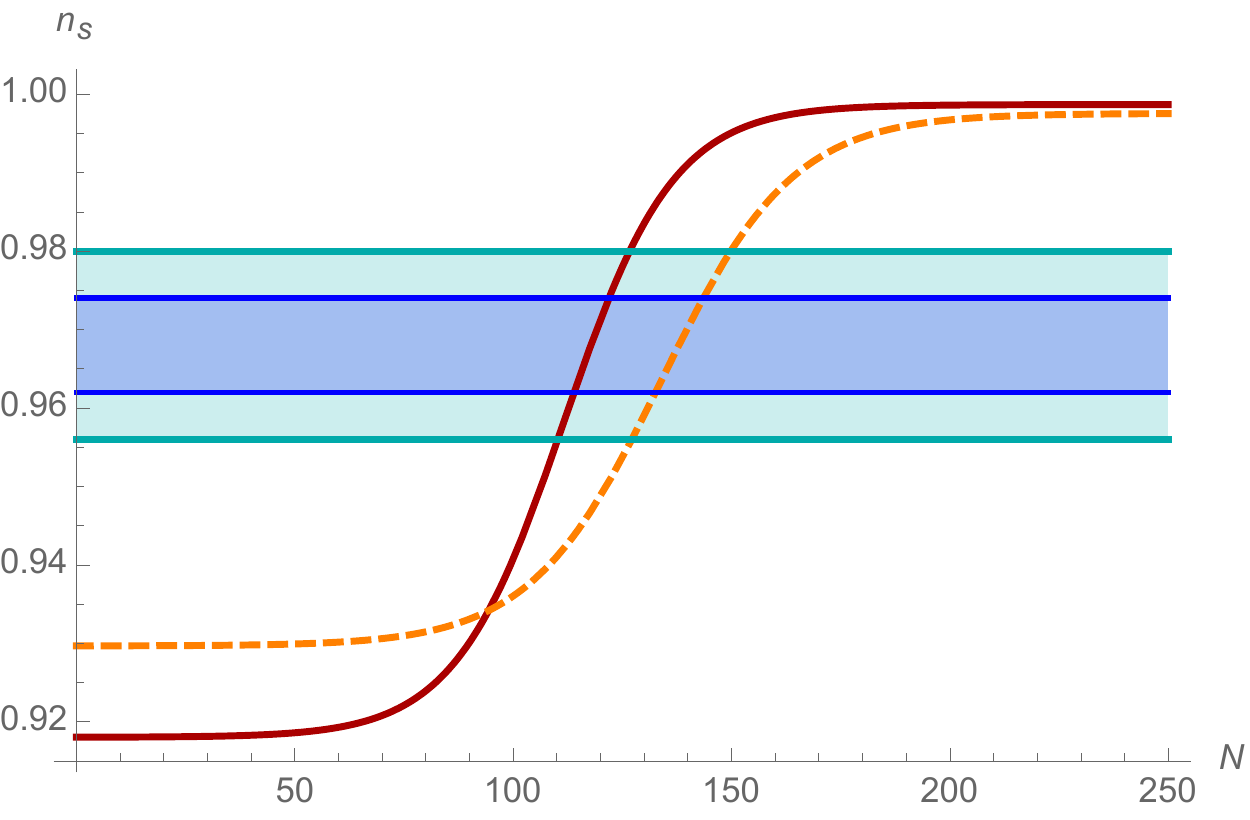}\
\caption{Evolution of the tilt with two different values for the slow-roll parameter $\eta_0$, the solid curve corresponding to $\eta_0=-1/25$ and the dashed to $\eta_0=-1/30$.  Different initial conditions and overall scale of the potential will shift the transition region to the left or right.  These curves illustrate that the transition takes a few dozen e-folds to pass through the observable window, a naive theorist's reconstruction of which, based off \cite{Ade:2015lrj}, is shown as the shaded region.}
\label{hillns}
\end{figure*}

The power spectrum is given by
\beq
P_k=\frac{V_0^2}{m^4}\frac{\sigma^2(\phi^2+\frac34\sigma^2)}{(\phi^2+\sigma^2)^2}.
\eeq
The previous requirement that $\sigma<\phi$ doubly suits us, because as it can be seen from above it also corresponds to the regime where $P_k\rightarrow 3/(4\eta_o^2)\gg 1$, certainly of no phenomenological relevance.  In the limit $\sigma\ll\phi$ the power spectrum becomes
\beq
P_k\rightarrow \frac{V_0^2}{m^4}\frac{\sigma^2}{\phi^2}
\eeq
where it is understood that for small values of the diakyon this expression should recover the standard result by saturating to the de Sitter temperature.  This form displays the scale invariant behavior uncovered in the general case, owing to the fact that both fields have the same exponential dependence on e-fold time (making this regime also of no phenomenological relevance).  If, instead, the diakyon starts at values much smaller than the de Sitter temperature, we recover the standard predictions
\beq
n_s-1=6\eta_0e^{2\eta_0(N_{\text{CMB}}-N)}+2\eta_0,\quad r=-16\eta_0e^{2\eta_0(N_{\text{CMB}}-N)}.
\eeq
The approximations we have used break down for $\eta_0\ll N_{\text{CMB}}^{-1}$, but in this case the amount of e-folds is so great that the last 60 will occur at the base of the potential, which is practically linear, and does not really resemble hilltop inflation anyway.  The interesting regime is when the diakyon starts smaller than the de Sitter temperature and then, through the course of its exponential growth, transitions to the regime of perfect scale invariance and negligible tensor-to-scalar ratio.  This transition actually takes place on the order of 50 to 100 e-folds, as shown in Fig \ref{hillns}, so that this is not a singular event that would require tuning to occur within the observable window.  This also ensures that the running of the power spectrum remains unobservably small throughout the evolution, in agreement with current observations.

We can comment on other versions of hilltop inflation before we conclude this section.  For the case of quartic hilltop inflation, which lies well within the best fit region in the Planck data, the spinodal effects produce
\be
V^\mathbb{H}_{\text{ quartic hilltop}}(\phi,\sigma)=V_0-\lambda\left(\phi^4+6\sigma^2\phi^2+3\sigma^4\right).
\ee
So that the diakyon direction is actually much steeper than the inflaton direction.  This necessitates the initial values to be even more fine tuned for this to match observations.

\subsection{Flattened Potentials}\label{flatinf}
\begin{figure*}[h]
\centering
\includegraphics[height=6.25cm]{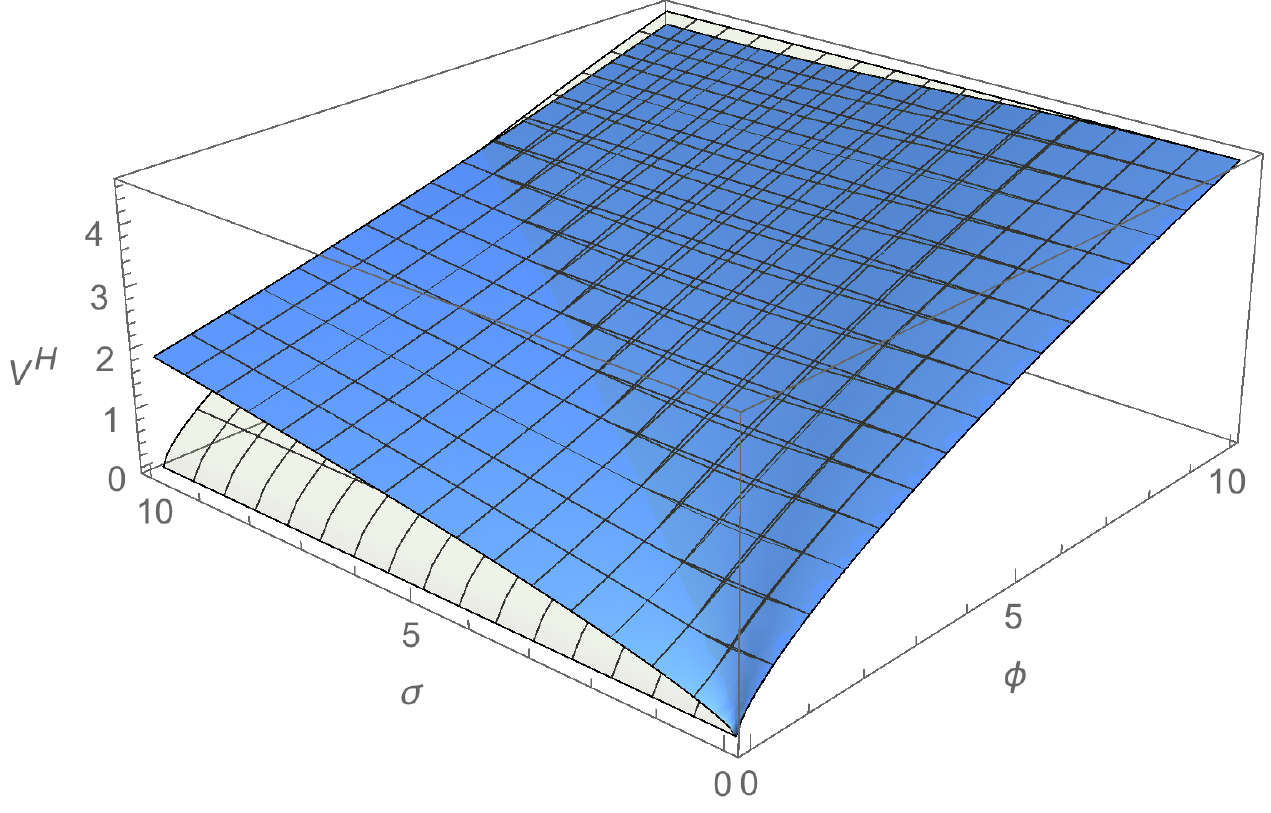}\
\caption{The full potential for $\phi$, $\sigma$, overlaid with the $\sigma$-independent original potential.  For small initial values of the diakyon the inflaton dynamics exactly corresponds to the usual case, and the diakyon experiences an instability towards larger field values.}
\label{two thirds}
\end{figure*}

There are certain contexts where integrating out heavy degrees of freedom serves to flatten the inflaton potential from an initially steeper profile for large field values \cite{Dong:2010in}.  This can cause the field to be driven by an effective potential that is a fractional power of the field.   A typical example of such a potential that will be testable in the coming years is
\be
V_{\text{flat}}(\phi)=\mu^{10/3}\phi^{2/3}.
\ee
The standard predictions for this potential are
\be
\phi(N)=\sqrt{\frac43(N_0-N)}M_p,\quad n_s-1=-\frac{4}{3N_{\text{CMB}}},\quad r=\frac{8}{3N_{\text{CMB}}},\label{stand23}
\ee
with $N_0$ being the beginning of inflation and $N_{\text{CMB}}$ the amount of enfolds since CMB scales left the horizon.  However, on this potential the field is tachyonic, and so we should expect a region of initial conditions where spinodal effects can become important.  To see this we apply the Hartree transform to this potential, yielding a result in terms of confluent hypergeometric functions
\be
V^\mathbb{H}_{\text{flat}}(\phi,\sigma)=\mu^{10/3}\left[\frac{\Gamma\left(5/6\right)\sigma^{2/3}}{2^{2/3}\pi^{1/2}}{}_1F_1\left(-\frac{1}{3};\frac{1}{2};-\frac{\phi^2}{2\sigma^2}\right)+\frac{\Gamma\left(4/3\right)\phi}{2^{1/6}\pi^{1/2}\sigma^{1/3}}{}_1F_1\left(\frac{1}{6};\frac{3}{2};-\frac{\phi^2}{2\sigma^2}\right)\right].
\ee
This shows the true utility of (\ref{hart}), allowing for an analytic expression for the full potential.  The latter is displayed in Fig.[\ref{two thirds}].

\begin{figure*}[h]
\centering
\includegraphics[height=5.25cm]{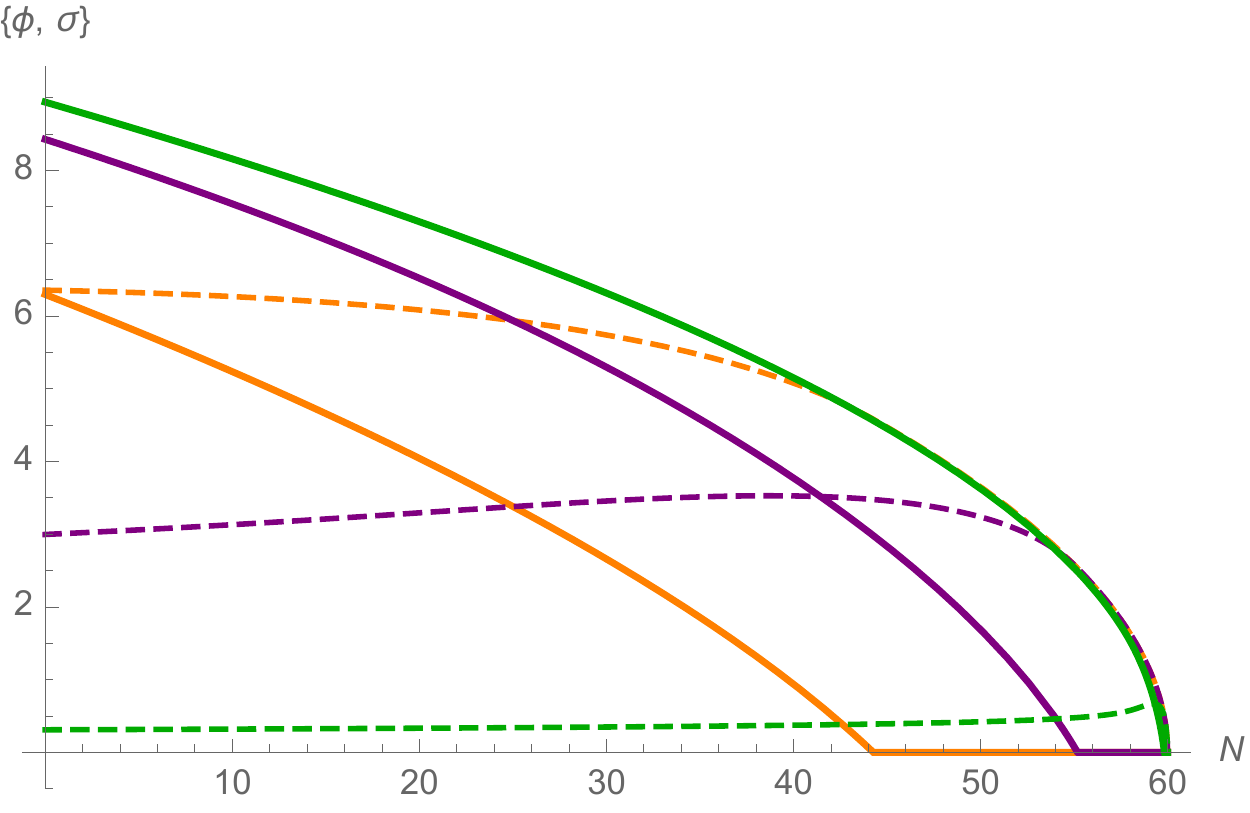}\
\caption{Inflationary trajectories during the last 60 e-folds for different initial field values.  The solid lines correspond to the inflaton, and the dashed lines correspond to the diakyon.  For small initial values of the diakyon, as in the green curves (solid top and dashed bottom), it never comes to dominate, and inflation proceeds as usual.  The purple (intermediate) and orange (coincident at start) curves exhibit extreme values of the diakyon, where it comes to dominate at late times, leading to a few extra e-folds of inflation.  These values lead to a very pathological amount of power, so should be treated as exaggerated scenarios demonstrating the effect present in smaller quantities for more modest values.}
\label{paths}
\end{figure*}

The previous expression is not very inviting of an analytical treatment, but it is easy enough to work with numerically, and in asymptotic limits.  For small values of the diakyon the original potential is recovered, as must be the case.  The dynamics of several choices of initial conditions are shown in Fig.[\ref{paths}].

However, for Planckian values of the diakyon, the amplitude of the power spectrum is well over unity, which is nowhere close to the observed value.  Therefore, we restrict our analysis to values $\sigma\ll M_p<\phi$, for which case the potential can be approximated as
\beq
V^\mathbb{H}(\phi,\sigma)\rightarrow\mu^{10/3}\left(\phi^{2/3}-\frac19\frac{\sigma^2}{\phi^{4/3}}\right).
\eeq
In the approximations we have made, the inflaton trajectory is barely altered from the standard result (\ref{stand23}), and the diakyon trajectory is
\beq
\sigma(N)=\frac{\sigma_0}{\left(1-\frac{N}{N_0}\right)^{1/6}}.
\eeq
This blows up at the end of inflation, but the approximations we have made break down well before that, as can be seen in Fig. \ref{paths}.  In the regime where $\sigma\gg\phi$, we can use the fact that ${}_1F_1(a;b;0)=0$ for all $a$, $b$ to find the limiting form of the potential
\beq
V^\mathbb{H}_{\text{flat}}(0,\sigma)\rightarrow\frac{\Gamma(5/6)}{2^{2/3}\sqrt{\pi}}\mu^{10/3}\sigma^{2/3},\label{.4}
\eeq
which is of the exact same form as the original potential, but with an additional numerical factor $\approx 0.4$.  Before this regime is reached, though, the diakyon grows with time, characteristic of tachyonic behavior, and directly opposed to the lowest order approximation, where it tracks the value of the potential which decreases with time.  This causes an interesting behavior: there is a discernible difference between inflation on this potential that comes from an eternal regime, and inflation that started only several e-folds outside of the observable window.  To highlight the difference between these two scenarios, let us assume that inflation began at the eternal regime $P_k\approx1$ \cite{Linde:1986fd}, with the initial value of the diakyon set by the de Sitter temperature (\ref{dstemp}):
\beq
\phi_0=\left(\frac49\right)^{3/8}\frac{M_p^{9/4}}{\mu^{5/4}},\quad V_0=\sqrt{\frac23}M_p^{3/2}\mu^{5/2},\quad \sigma_0^2=\frac{1}{12\pi^2}\sqrt{\frac23}\frac{\mu^{5/2}}{M_p^{1/2}}.
\eeq
For simplicity we use the approximation that the fields start on their slow-roll attractor behavior, with negligible initial inhomogeneities.  The total amount of e-folds from this regime is $N_e=(M_p/\mu)^{5/2}/\sqrt{6}$, much larger than the observed 60 or so e-folds.  By the time the fields have evolved down to CMB scales, the value of the diakyon is 
\beq
\sigma_{\text{CMB}}^2=\frac{1}{12\pi^2}\frac{2^{1/3}}{3^{2/3}}\frac{1}{N_{\text{CMB}}^{1/3}}\mu^{5/3}M_p^{1/3}=\frac{1}{6^{1/3}}\frac{1}{N_{\text{CMB}}^{2/3}}\left(\frac{M_p}{\mu}\right)^{5/3}\left(\frac{H}{2\pi}\right)^2\gg \left(\frac{H}{2\pi}\right)^2.
\eeq
The diakyon is still subplanckian, so our expansion is valid, but it is much larger than its vacuum value.  In this regime the fluctuations behave purely spinodally.  Normalization of the power spectrum at $N_{\text{CMB}}=60$ then requires $\mu=1.6\times10^{-5}M_p$, and the predictions for the model become
\beq
n_s=1-\frac{2}{3N_{\text{CMB}}}= 0.989,\quad r=\frac{7.5\times10^{-7}}{N_{\text{CMB}}}=1.3\times10^{-8},
\eeq
which are clearly different from (\ref{stand23}), with half the predicted tilt and practically vanishing tensor-to-scalar ratio.  These values of the parameters actually lie substantially outside of the observed values, making this regime in disagreement with experiment.  

There is a general trend that is indicative of all tachyonic models with an eternal regime:  if inflation proceeds for a sufficiently long time, the diakyon takes over.  Contrary to the usual predictions for large field models in this class, we would not expect to see tensor modes, and if application of the usual expression for the tilt were applied, it would lead us to the wrong conclusion about the parameters and shape of the potential.  If we were confident that inflation took place on a given potential, however, observations would allow us to infer whether inflation lasted parametrically longer than the observed amount by measuring the amount of deviations from the standard predictions.

We would like to conclude our analysis of flattened potentials with a brief remark on another scenario:  the Starobinsky ``$R+R^2$'' model \cite{Starobinsky:1980te}.  This analysis is particularly relevant here since it is a small field model, so if spinodal effects are present the spectrum looks scale invariant.  If we assume that the diakyon starts at the de Sitter temperature, then straightforward application of (\ref{integ}) yields
\beq
\sigma_{\text{CMB}}=\left(\frac{N_{\text{tot}}}{N_{\text{CMB}}}\right)^2\left(\frac{H}{2\pi}\right)^2
\eeq
Hence, if inflation lasted any longer than the minimum required to solve cosmological problems, the backreaction due to the diakyon would be important and engender a shift in $n_s$ towards exact scale invariance.

\subsection{Monodromy Inflation}\label{monoinf}

As a final application of the Hartree transform we consider axion monodromy inflation \cite{Silverstein:2008sg,McAllister:2008hb}.  In the simplest realization of this scenario, inflation is driven by an axion field in which stringy effects break the exact periodicity of the potential.  This leads to an approximately linear potential with small modulations,
\be
V_{\text{mono}}(\phi)=\mu^3\phi+\Lambda^4\cos\left(\frac{\phi}{f}\right).\label{monopot}
\ee
Inflation is driven by the linear term, yielding standard predictions for inflationary parameters: $n_s-1=-3/(2N_{\text{CMB}})$ and $r = 4/N_{\text{CMB}}$.  In addition, the sinusoidal correction leads to characteristic oscillations in the power spectrum that are logarithmic in scale \cite{Flauger:2009ab} and have yet to be found at a significant level \cite{Easther:2013kla}, but can be arbitrarily small depending on the parameter $\Lambda$.  One is free to take natural values of the axion decay constant $f\sim \mathcal{O}(\frac{1}{10}-\frac{1}{100})M_p$, in which case the field undergoes many oscillations in a single e-fold, or the cycles could be so large to encompass the entire range of observed CMB scales, possibly providing a running spectrum that explains lack of power both on large scales and in the matter power spectrum \cite{Minor:2014xla}.  

Monodromy inflation is a large field model, with 
\be
\phi(N) = \sqrt{2(N_{\text{CMB}}-N)}M_p.\label{phiofn}
\ee
If spinodal effects are not too large, the normalization of the spectrum enforces $\mu = 6 \times 10^{-4} M_p$.  The other parameters can be expressed in a convenient combination $b = \Lambda^4/(\mu^3f)$.  If this parameter is too large, $b > 1$, the potential will not be monotonically decreasing, but will instead develop pockets that, if too deep, could potentially cause the inflaton to become trapped in a false minimum instead of undergoing slow roll. 

\begin{figure*}[h]
\centering
\includegraphics[height=6.25cm]{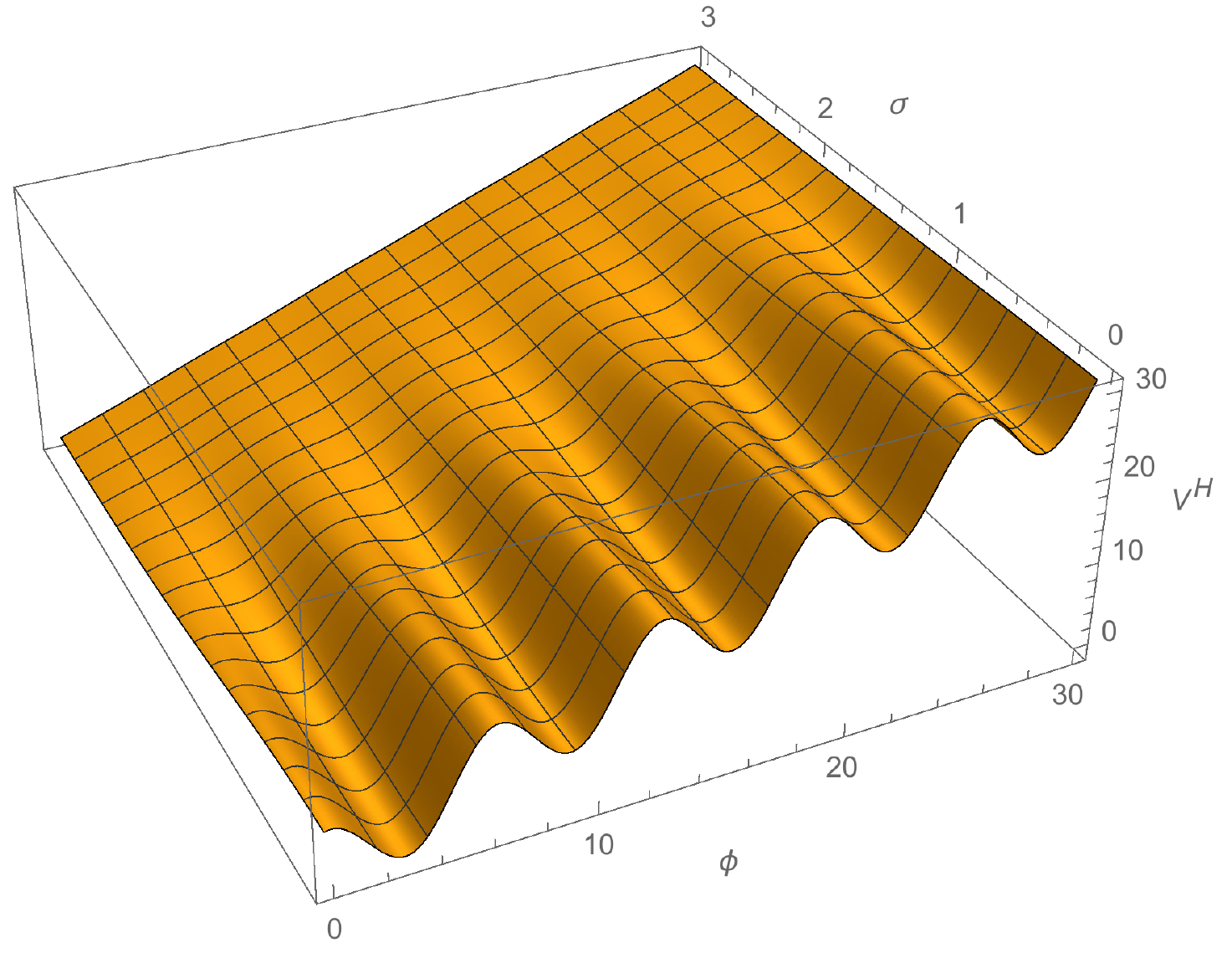}\
\caption{The two-field potential.  For large values of the diakyon the oscillations are washed out, and the inflaton direction is strictly linear.}
\label{mono}
\end{figure*}

We would like to investigate to what extent spinodal effects might change this story.  Taking the Hartree transform of the potential (\ref{monopot}) yields
\be
V^\mathbb{H}_{\text{mono}}(\phi,\sigma)=\mu^3\phi+\Lambda^4\cos(\phi/f)e^{-\sigma^2/2f^2}.
\ee
The linear term remains untouched, and the cosine, as an eigenfunction of the Hartree transform, is now multiplied by a gaussian.  The potential is plotted in Fig [\ref{mono}].

We see that allowing for tachyonic modes to build up and backreact on the background evolution can serve to taper the size of the oscillations, depending on the initial conditions taken for the inhomogeneity.  An important point is that spinodal effects prevent the usual restriction to an exactly homogeneous slow-roll evolution from being an attractor, so that only for a narrow range of field space $\sigma\lesssim f$ will these oscillations occur at all.  For generic values of the diakyon, the potential will look exactly linear, with no additional signals.  As we have argued in section \ref{Hart}, if inflation is to have come from an eternal regime, then we would expect a much larger initial value of the diakyon field.  We note that spinodal effects can provide a route for slow-roll inflation to occur even for large values of $b$, which would have been traditionally excluded.

The line $\sigma=0$ corresponds to the traditional slow-roll path, which gives the strongest possible oscillation signal.  From the form of the potential one can expect that during the periods when the inflaton $\phi$ is near the top of a hill the fluctuations roll towards the plateau region, corresponding to the buildup of tachyonic-wavelength modes.  Before this effect can get too strong, however, the inflaton goes through a valley region, at which point the diakyon tends towards decreasing values.  If the starting point is well out on the tail of the Gaussian, the diakyon will barely change its value at all, and the evolution of the inflaton will look effectively linear.  These qualitative effects can be encapsulated in an analytical approximation below.

To get a stronger understanding of how the diakyon behaves during inflation, we make a few simplifying assumptions to make its time dependence analytically solvable in terms of special functions.  We first take the time dependence of the inflaton to be given by its unperturbed slow-roll trajectory (\ref{phiofn}), which is valid to lowest order in slow roll and $b$.  Secondly, we take $\sigma$ to be in slow roll as well, which is justified because the slope of the Gaussian is $\frac{\partial V}{\partial\sigma}\lesssim b\frac{\partial V}{\partial\phi}$, and so is comparatively shallow to the inflaton direction unless $b\gg 1$.  Third, if the Hubble rate is dominated by the linear term in the potential, then we can write
\be
\frac{\mu^3}{M_p}\sqrt{2(N_0-N)}\frac{d\sigma}{dN}=\frac{\Lambda^4}{f^2}\cos\left(\frac{M_p}{f}\sqrt{2(N_0-N)}\right)\sigma e^{-\sigma^2/2f^2},
\ee
which we can integrate to get the following expression for $\sigma(N)$ in terms of the exponential integral $\text{Ei}(x)$:
\be
\text{Ei}\left(\frac{\sigma(N)^2}{2f^2}\right)-\text{Ei}\left(\frac{\sigma_0^2}{2f^2}\right)=-2 b\sin\left(\frac{M_p}{f}\sqrt{2(N_0-N)}\right).\label{exactsigma}
\ee

\begin{figure*}
\centering
\includegraphics[height=5.25cm]{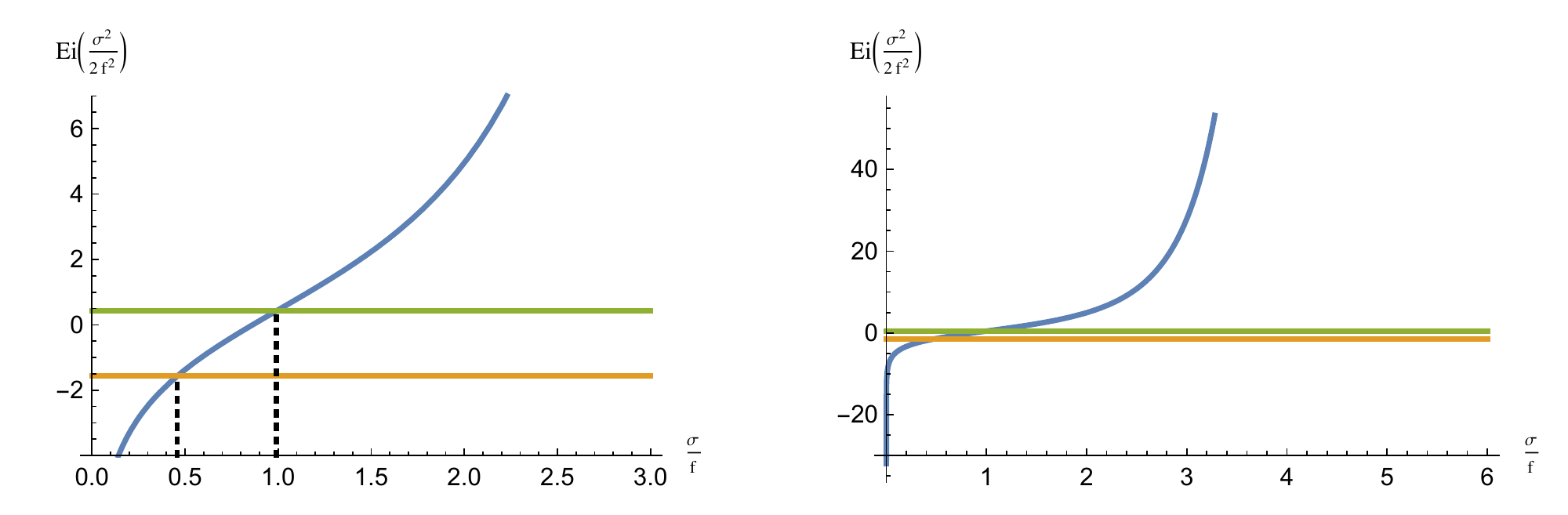}\
\caption{Implicit solution to equation (\ref{exactsigma}) for example values $b=0.5$ and $\sigma_0=0.7f$.  The diakyon oscillates between the two dashed curves during each cycle.  Increasing the initial value of the diakyon field corresponds to shifting the horizontal lines upwards, while increasing $b$ corresponds to increasing the vertical width between them.  Both for small and large initial values, there is hardly any change in the field at all.}
\label{expei}
\end{figure*}

From this expression we can remark that the field oscillates between two extremes corresponding to the maxima and minima of the sine function, so that one full cycle brings the $\sigma$-field back to where it originally started.  Moreover, the amplitude of this oscillation cycle does not change, and there is no long-term drift of the central value.  This curve, along with extremum values for the field, is plotted in Fig. [\ref{expei}] for example values of the parameters.

Notice that the oscillation can be quite lopsided.  If we increase the value of $b$, corresponding to more pronounced hills and valleys, the range of $\sigma$ will be greater.  For very small initial values, the maximum value of $\sigma$ is negligible, and spinodal effects can be ignored.  This corresponds to placing the trajectory at the very top of the hill, where there is no strong slope to provide a force away.  Very large values of the field are plotted on the right plot, for which the field does not move very much at all, regardless of the value of $b$.  In this regime asymptotic limits can be made in (\ref{exactsigma}) to find the total range of $\sigma$ during one cycle to be
\be
\Delta\sigma\rightarrow b\sigma_0e^{-\sigma^2/2f^2}.
\ee 
This corresponds to regions far enough out on the Gaussian tail that the potential looks exactly flat, and so 
the diakyon is roughly stationary.

We now discuss the parameter values for which the diakyon can effectively screen the small scale perturbations.  If the fluctuations in (\ref{spectro}) are dominated by the potential and not the diakyon, we require $\sigma^2\ll V/M_p^2$, which translates into $\sigma\leq 4.5\times10^{-6}M_p$.  In the event that the diakyon dominates the fluctuations, which will occur if the scale $\mu$ is smaller than the standard value, correct normalization enforces that this equality be saturated.  In this case the oscillating value of $\sigma$ can reintroduce oscillations in the power spectrum, but these will be large only if $\sigma\approx f$, which is an apparently fine-tuned scenario that we disregard.  Screening becomes efficient for $\sigma\gtrsim4f$, which means that the mechanism we have described can only be operational for small values of $f$, say $f<10^{-6}M_p$.  This value may seem somewhat low from the perspective of high energy physics models, but corresponds to the upper value of the ``classic window'' for axion dark matter \cite{Hertzberg:2008wr}. Monodromy models with different powers of the inflaton are sometimes considered \cite{McAllister:2014mpa}. Taking into account these models does not substantially raise the value of $f$ needed for the diakyon's screening of small scale perturbations to take place. However, the dynamics may be more interesting, as the mass of the inflation causes the diakyon to increase (or decrease) with the evolution, which could cause the oscillations to appear only at early (or late) times.

\section{Conclusions}\label{conclusions}
\begin{figure*}[h]
\centering
\includegraphics[height=6.25cm]{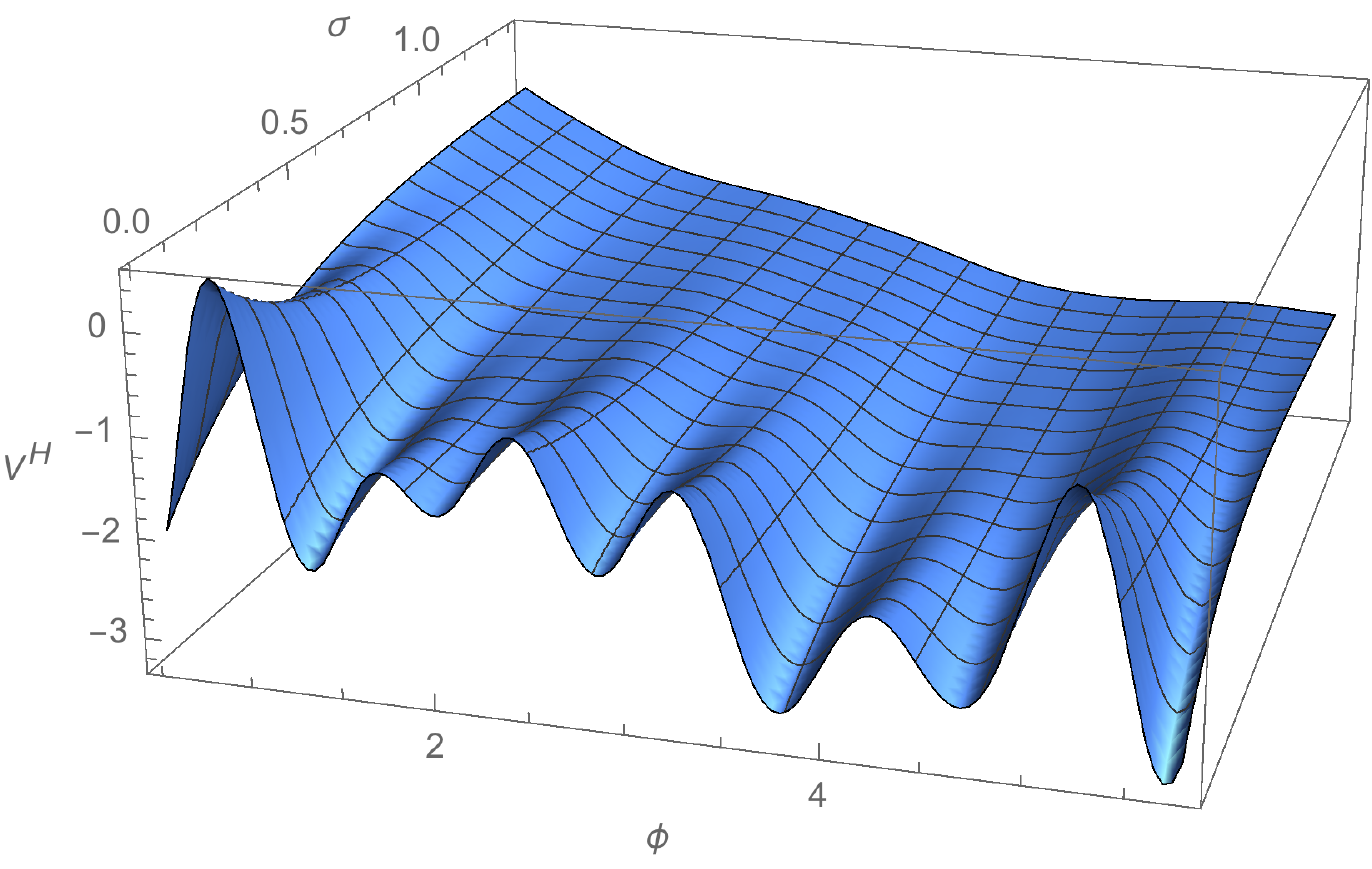}\
\caption{Spinodal effects in a generic random potential.  For small values of the diakyon the shown potential is completely unsuitable for slow-roll inflation, containing many false minima and steep peaks.  Backreactions tame these features by smearing over the original potential with a Gaussian filter.  This picks out the large scale trend of the potential, providing a route for slow-roll inflation to proceed.}
\label{pocketses}
\end{figure*}

We have considered the influence of backreaction from inflationary fluctuations for a few typical models of inflation.  We showed that the strength of the influence is largely dependent on the initial conditions of the inflaton, and whether the dynamics are tachyonic enough to drive the diakyon to large field values.  If the long wavelength modes are initially populated close to the de Sitter equilibrium values, backreaction is negligible and inflation proceeds as normal, unless a very long time elapses since the beginning of inflation.  If the modes are initially populated to a much greater extent, then it can significantly alter the inflationary dynamics and predictions, by influencing the strength of the generated perturbations.  Though the details of inflationary dynamics can be substantially modified, the basic predictions of the scenario, mainly a stage of near exponential expansion, generating a nearly scale invariant spectrum of perturbations, remains the case.  We note four general trends.  The predicted value of the spectral tilt deviates slightly from the original value in large field models, and moves strongly towards scale invariance in small field models.  Secondly, since the size of the fluctuations is no longer set by the energy scale of the potential during inflation, inflation can proceed at a much lower scale and still produce the observed size of fluctuations.  This causes the tensor-to-scalar ratio to drastically decrease.  Thirdly, the eventual tendency towards the tachyonic regime allows us to infer the duration of inflation before observable scales.  Lastly, if the potential has any sort of features, the strong backreaction regime serves to smear out these features, providing a smoother potential for the inflaton.  This can occur even for features so deep that they would have prevented the inflaton from reaching its true minimum at all.  This was studied in the simple case of monodromy inflation, but holds in generic potentials, as displayed in Fig. \ref{pocketses}.  

We remark that this washing out of features is not enough to completely solve the $\eta$ problem, which is the question of how the mass of the inflaton can be lighter than the Hubble scale, even after quantum corrections are taken into account.  This stems from the fact that the size of the features that are washed out depends on the value of the diakyon.  If we Fourier expand a potential, $V(\phi)=\int dn V_n \cos(n\phi/M_p)$, the Hartree transformed potential multiplies each coefficient by a gaussian $e^{-n^2\sigma^2/M_p^2}$, which effectively cuts off the value of the integral at $n\approx M_p/\sigma$.  If all coefficients are of the same order, then this yields $\eta\approx M_p^3/\sigma^3$, but since we require $\sigma\lesssim\sqrt{\epsilon P_k}M_p$ we arrive at $\eta\gtrsim 10^{18}\gg1$.  The only way to circumvent this conclusion is if the coefficients of the Fourier series become tiny for suitable values of $n$, but this requires working on a potential where the $\eta$ problem was solved originally.  As it stands, this mechanism is only suitable for removing features that are much smaller than the typical field range.

The complicated nature of the strong backreaction regime forced us to employ the Hartree method to encapsulate the qualitative aspects of these effects.  We expect it to be a reasonable estimate of the dynamics based on its success in other scenarios.  The overall effect of this approximation is to replace the original single field dynamics by another field that governs the strength of the fluctuations.  It is important to make an estimate of the next order corrections to this approximation to ensure that the calculations we performed are under full control.  The path integral approach to the Hartree approximation is a simple way to do this, and we plan to return to this in future publications.  Nevertheless, we expect the qualitative features we have uncovered to remain true in the full theory.

There are a number of other issues worth exploring.  We employed the Hartree approximation in the uniform curvature gauge, where all perturbations were in the inflaton field.  In general, there is no guarantee that resummation schemes are gauge invariant, and so it would be interesting to extend this framework to gauge invariant perturbation theory.  The Hartree approximation is also known to not respect global symmetries \cite{Pilaftsis:2013xna}, and in the simplest settings provides spurious masses to Nambu-Goldstone bosons of broken symmetries.  As the inflaton can be thought of the Nambu-Goldstone boson of time translation invariance \cite{Cheung:2007st}, it might be worth considering if a modified version of the Hartree approximation such as used in \cite{Pilaftsis:2013xna} is more appropriate.  We have also only considered strong backreaction effects on the scalar two-point correlator.  It might be interesting to examine if its effects on the tensor power spectrum can be similar, and what effect it might have on nongaussianities.  The latter consideration might even necessitate a generalization of the above scenario to a 3-particle irreducible formalism, similarly causing the strength of nongaussianities to be dependent on initial conditions.  We leave these as questions for the future.

\section*{Acknowledgements}

\smallskip
We would like to thank Rich Holman for guidance and comments on the initial version of this draft. Additionally, B.R. would like to thank Andy Albrecht for useful discussions. B.R. was supported in part by DOE grant DE-FG03-91ER40674. 

\bibliographystyle{unsrt}
\bibliography{modernSI}

 

\end{document}